\documentclass[twocolumn,english]{revtex4}
\usepackage[T1]{fontenc}
\usepackage{color}
\usepackage{amsmath}
\usepackage{graphicx,epsf}
\usepackage{esint}
\usepackage{subfig}
\usepackage{mathrsfs}
\usepackage{epstopdf}
\usepackage{babel}
\makeatletter

\begin{document}

\title{Drift wave soliton formation via forced-driven zonal flow and implication
on plasma confinement}

\author{}
\author{Ningfei Chen$^1$, Liu Chen$^{1,2}$, Fulvio Zonca$^{2,1}$ and Zhiyong Qiu$^{1,2,}$\footnote{Author to whom correspondence should be addressed: zqiu@ipp.ac.cn}}

\affiliation{$^1$Institute for  Fusion Theory and Simulation, School of Physics, Zhejiang University, Hangzhou, P.R.C\\
$^2$Center for Nonlinear Plasma Science and   C.R. ENEA Frascati, C.P. 65, 00044 Frascati, Italy}

\begin{abstract}
In this work, gyrokinetic theory  of drift waves (DWs) self-regulation via the forced driven
 zonal flow (ZF) is presented, and finite diamagnetic drift frequency due to plasma nonuniformity  is shown to  play dominant  role in ZF forced generation.   The obtained nonlinear DW equation  is a nonlinear Schr\"odinger equation, in which the
linear dispersiveness, linear growth,  nonuniformity of diamagnetic drift frequency,  and cubic
nonlinearity induced by   feedback of forced-driven ZF to DWs are self-consistently
included.  The nonlinear DW equation  is solved numerically  in both uniform and nonuniform
plasmas. It is shown that DW envelope soliton may form due to the balance of linear dispersiveness and nonlinearity, and lead to turbulence spreading to linearly stable region. It is further   found that though the threshold on DW amplitude for
soliton formation is well within the relevant parameter regimes of realistic tokamak experiments, solitons
can not extend beyond the range bounded by the turning points of the
wave packet when plasma nonuniformity is self-consistently accounted for.
\end{abstract}
\maketitle

\section{Introduction}

Understanding the triggering and regulation mechanisms of anomalous
transport is a significant issue for magnetically confined fusion research.
Drift waves (DWs) turbulence \cite{WHortonRMP1999}, driven by  free energy associated with plasma
pressure nonuniformities  intrinsic to confined plasmas, are considered as
important candidates  for inducing anomalous transport. As a consequence,
 understanding the nonlinear dynamics of DWs, including regulation, saturation and spreading to linearly stable region, is crucial for assessing  the confinement of plasmas.
Numerical simulations \cite{ZLinScience1998,XLiaoPoP2016,ADimitsPoP2000} and experiments
\cite{KZhaoPRL2006,GConwayPPCF2008,TLanPoP2008,WZhongNF2015,AMelnikovNF2017}
have found that zonal flow (ZF) \cite{LChenPoP2000,PDiamondPPCF2005} generated by DWs can significantly
reduce turbulence amplitude and the associated transport. Previous analytical
theories on ZF generation by DWs have been   focused on the spontaneous
excitation via modulational instability \cite{LChenPoP2000,PGuzdarPoP2001}; in which
DWs are scattered into linearly stable short radial wavelength domain
by the nonlinearly excited ZF \cite{LChenPoP2000,KHallatschekPRL2001}. This nonlinear process has a  finite threshold on DWs amplitude
determined by the frequency mismatch, and the ZF growth rate, meanwhile,
scales with the amplitude of DWs. However, it is very often found
in numerical simulations that ZF grows at twice the DW  instantaneous
linear growth rate \cite{HWangPoP2020,GDongPoP2019}; suggesting ZF generation via the so called  ``forced-driven'' process  \cite{ZQiuPoP2016} or ``passive-excitation'' \cite{GDongPoP2019}.
Though forced-driven ZF by electromagnetic Alfv\'en waves has been extensively   investigated
  \cite{YTodoNF2010,ZQiuPoP2016,ABiancalaniPPCF2021}, theoretical interpretation on forced-driven ZF by electrostatic DWs has not been revealed up to now.

In addition, turbulence spreading from linearly unstable region of
DWs to the linearly stable region is important for the plasma confinement \cite{XGarbetNF1994,TSHahmPPCF2004,HahmPoP2005,FZoncaPoP2004},
because it might lead to the nonlocality of turbulence transport,
which further results in the change of transport size scaling \cite{ZLinPRL2002}.
In the existing theoretical studies, DW  solitons, formed due to the
spontaneously excited ZF \cite{ZGuoPRL2009} and/or its finite frequency
counterpart geodesic acoustic mode (GAM) \cite{NWinsorPoF1968,FZoncaEPL2008,ZQiuPST2018,GConwayNF2021,NChenPPCF2022}, are
found to contribute to turbulence spreading, because DW turbulence can be radially trapped by ZFs \cite{PDiamondPPCF2005,FZoncaPoP2004,ZGuoPRL2009} and solitons are
characterized by preserving their amplitude and shape during propagation.
Coherent DW  solitons are formed as the linear dispersiveness of DWs
is balanced by nonlinear wave trapping effect induced by ZF generated via
modulational instability. As a consequence,
ZF generated via forced-driven process  could also be expected  to generate DW  solitons
and enhance turbulence spreading in a similar way.

In this work, a paradigm model of DWs self-beating, i.e., forced-driven
process, to generate zero frequency ZF is derived using nonlinear
gyrokinetic theory. The obtained nonlinear DW equation is a nonlinear
Schr\"odinger equation (NLSE), in which the linear dispersiveness, linear
growth rate, plasma nonuniformity and cubic nonlinearity induced by
feedback of forced-driven ZF to DW are self-consistently included.
The NLSE is systematically investigated in uniform and nonuniform
plasmas, with emphasis on the generation and propagation of solitons,
and its consequences on plasma confinement. In uniform plasmas, soliton
structures are formed, by balancing the linear dispersiveness and
cubic nonlinearity, after DW amplitude reaching certain threshold;
and, thereby, leading to enhanced turbulence spreading. It is found
that the threshold on DW amplitude for soliton formation is $e\delta\phi/T_{{\rm i}}\simeq0.02$,
which is within the experimentally relevant parameter regime.   Here, $e$ is the unit charge, $T_{\rm i}$ is the ion temperature, and $\delta\phi$ is the perturbed DW scalar potential. In nonuniform
plasmas, the evolution of the corresponding DW eigenstates are investigated.
It is found that the extent for wave propagation is not sensitive to either the existence or strength of nonlinearity, which implies that in nonuniform plasmas, solitons
can not extend beyond the range bounded by the turning points of the
wave packet induced by the nonuniformity of diamagnetic drift frequency. Analytic theory found that in
the reference   frame of the wavepacket, the envelope follows the same
NLSE as that in uniform plasmas and the trajectory of envelope is essentially
determined by the nonuniformity.

The rest of the paper are organised as follows: in Sec. \ref{sec:Theoretical-model},
the general nonlinear gyrokinetic framework is presented. Based on the theoretical model, the forced-driven ZF by DWs and feedback of ZF to DWs are investigated, respectively, in Sec. \ref{sec:forced_driven} and \ref{sec:feedback}, which finally yields a NLSE governing the nonlinear evolution  of DW. It is  then  investigated
in both uniform and nonuniform plasmas. The soliton generation and
propagation in uniform plasmas are investigated in Sec. \ref{sec:Uniform-plasmas};
while, in Sec. \ref{sec:nonuniform-plasmas}, the evolution of the corresponding
linear eigenstates in nonuniform plasmas  is considered. Finally, conclusion and discussions are provided in Sec. \ref{sec:Conclusion-and-discussion}.

\section{Theoretical model\label{sec:Theoretical-model}}

In this work , we consider the forced-driven process, in which a single-$n$ DW $\varOmega_n(\omega_{n},k_{\theta n})$ couples with its complex conjugate
$\varOmega^*_n(-\omega_{n},-k_{\theta n})$ to generate ZF $\varOmega_{\rm Z}(\omega_{{\rm Z}},k_{r{\rm Z}})$
with $\omega_{{\rm Z}}\approx0$, $k_{\parallel{\rm Z}}=0$, $k_{\theta{\rm Z}}=0$ and finite $k_{r{\rm Z}}$. Here, $n$/$m$ are the toroidal/poloidal mode numbers, $k_r$/$k_\theta$/$k_\parallel$ are the radial/poloidal/parallel wave numbers of modes, and the subscripts $n$ and $Z$ denote quantities associated with single-$n$ DW and ZF, respectively.
To simplify the analysis without loss of generality,  we assume that DW  and ZF are both electrostatic
fluctuations and $\gamma_{{\rm L}}/\omega_n\ll1$,
so that the parallel mode structure of DW  is not affected by nonlinear radial envelope modulation
process. Here, $\gamma_{\rm L}$ is the DW linear growth rate. In this case, the fluctuations can be expressed as
\begin{eqnarray}
\delta\phi_{n} & = & A_{n}\left(r,t\right){\rm e}^{{\rm i}n\zeta-{\rm i}\omega_{n}t}\sum_{m}{\rm e}^{-{\rm i}m\theta}\varPhi\left(nq-m\right)+c.c.,\\
\delta\phi_{{\rm Z}} & = & \varPhi_{{\rm Z}}{\rm e}^{{\rm i}k_{r{\rm Z}}r-{\rm i}\omega_{{\rm Z}}t}+c.c.,\label{eq:ZFZF-potential}
\end{eqnarray}
where $\varPhi(nq-m)$ is the fine radial structure due to finite
$k_{\parallel}$. For simplicity, the large aspect ratio tokamak with concentric
circular magnetic surfaces is considered, in which the magnetic field
is given by $\boldsymbol{B}=B_{0}[\boldsymbol{e}_{\zeta}/(1+\epsilon\cos\theta)+\epsilon\boldsymbol{e}_{\theta}/q]$,
where $\epsilon\equiv r/R\ll1$ is the inverse aspect ratio, $R$
and $r$ are major and minor radii of the tokamak, respectively, $\zeta$
and $\theta$ are the toroidal and poloidal angles, and $q$ is the safety factor. Nonlinear interactions among DWs and ZF can be investigated
using the electrostatic nonlinear gyrokinetic equation \cite{EFriemanPoF1982}
\begin{eqnarray}
 & \left(\omega+{\rm i}v_{\parallel}\partial_{l}+\omega_{{\rm Ds}}\right)\delta H_{k,{\rm s}}=\dfrac{q_{{\rm s}}}{T_{{\rm s}}}\left(\omega-\omega_{*{\rm s}}\right){\rm J}_{k}\delta\phi_{k}F_{0{\rm s}}\nonumber \\
 & -{\rm i}\dfrac{c}{B_{0}}\varLambda_{k',k''}^{k}{\rm J}_{k'}{\rm \delta}\phi_{k'}{\rm \delta}H_{k'',{\rm s}}.\label{eq:nl-gke}
\end{eqnarray}
Here, $\omega_{*{\rm s}}\equiv k_{\theta}cT_{{\rm s}}/(eBL_{n})$
is the diamagnetic frequency due to plasma density nonuniformity, where $L_{n}\equiv -N/(\partial N/\partial r)$ is the characteristic
length of ion density variation, $N$ is the equilibrium particle density, $\omega_{{\rm Ds}}\equiv\hat{\omega}_{{\rm ds}}C$
represents the magnetic drift motion, with  $\hat{\omega}_{{\rm ds}}\equiv\omega_{{\rm ds}}\left(v_{\perp}^{2}/2+v_{\parallel}^{2}\right) /v^2_{\rm ts}$,
$C\equiv\cos\theta-\sin\theta k_{r}/k_{\theta}$, $\omega_{\rm ds}\equiv k_\theta cT_{\rm s}/(eBR)$,  and $v_{{\rm ts}}\equiv\sqrt{2T_{{\rm s}}/m_{{\rm s}}}$ is the thermal
velocity.  ${\rm J}_k\equiv{\rm J}_{k}\left(k_{\perp}\rho_s\right)$
is the Bessel function of zero index describing finite Larmor radius (FLR) effect, $\rho_{{\rm s}}\equiv v_\perp/\omega_{{\rm cs}}$ is the Larmor radius,
 and $\omega_{{\rm cs}}\equiv eB/\left(m_{{\rm s}}c\right)$ is the cyclotron frequency. The subscript $s$
reperesents the particle species $s=i,e$.  For the clarity of the physics picture, the electron DW driven by plasma density nonuniformity is assumed, while the effects of temperature nonuniformity is neglected.    The nonadiabatic gyro-center response $\delta H_{k,{\rm s}}$
 can be separated
into linear and nonlinear components, i.e., $\delta H\equiv\delta H^{{\rm L}}+\delta H^{{\rm NL}}$,
with $\delta H^{{\rm NL}}\ll\delta H^{{\rm L}}$ and the subscript  $k$
representing quantities associated with the mode $\varOmega_k$. The second term
on the right hand side of Eq. (\ref{eq:nl-gke}) is the formally perpendicular nonlinear term, with $\varLambda_{k',k''}^{k}\equiv\sum_{\boldsymbol{k}=\boldsymbol{k^{'}}+\boldsymbol{k^{''}}}\boldsymbol{b}\cdot\left(\boldsymbol{k''}\times\boldsymbol{k'}\right)$
representing the selection rule on frequency and wavenumber matching conditions for mode-mode coupling, while other notations are standard. The nonlinear
gyrokinetic equation can be closed by the charge quasi-neutrality
condition
\begin{eqnarray}
\dfrac{e^{2}N}{T_{{\rm i}}}\left(1+\dfrac{1}{\tau}\right)\delta\phi_{k} & = & \left\langle e{\rm J_{0}}\delta H_{{\rm i}}\right\rangle _{k}-\left\langle e\delta H_{{\rm e}}\right\rangle _{k}.\label{eq:QN-condition}
\end{eqnarray}
Here, $\tau\equiv T_{{\rm e}}/T_{{\rm i}}$, and $\left\langle \cdots\right\rangle$
represents velocity space integration. The gyrokinetic theoretical framework will be used to derive the nonlinear equations for ZF forced excitation by DW, as well as the DW nonlinear evolution due to forced-driven ZF regulation.

\section{ZF forced driven by DWs}\label{sec:forced_driven}

In this section, the nonlinear generation of ZF by DWs self-beating is investigated.
The nonlinear gyrokinetic equation describing particle responses of
ZF, $\varOmega_{\rm Z}(\omega_{\rm Z},k_{r{\rm Z}})$, can be written as
\begin{eqnarray}
 & \left(\omega+{\rm i}\omega_{{\rm tr}}{\rm \partial}_{\theta}-\omega_{{\rm Drs}}\right)\delta H_{{\rm Z,s}}=\dfrac{q_{{\rm s}}}{T_{{\rm s}}}\omega{\rm J}_{{\rm Z}}{\rm \delta}\phi_{{\rm Z}}F_{0{\rm s}}\nonumber \\
 & -{\rm i}\dfrac{c}{B_{0}}\varLambda_{k',k''}^{k}{\rm J}_{k'}\delta\phi_{k'}\delta H_{k'',{\rm s}},\label{eq:gke-zf}
\end{eqnarray}
where $\omega_{{\rm tr}}\equiv v_{\parallel}/(qR)$ is the transit
frequency, $\omega_{{\rm Drs}}\equiv\hat{\omega}_{{\rm drs}}\sin\theta$,
and $\hat{\omega}_{{\rm drs}}\equiv k_{r}cT_{\rm s}(v_\perp^2/2+v_\parallel^2)/(eBR v^2_{\rm ts})$.

For electrons  with $v_{{\rm te}}/(qR)\gg\omega_n$,
 the electron response of DW is adiabatic, i.e., $\delta H_{n{\rm ,e}}=0$. Consequently,
  the nonlinear electron response for ZF $\delta H_{{\rm Z,e}}^{{\rm NL}}$
vanishes, as the result of vanishing source term in the nonlinear
gyrokinetic equation. Then, the electron responses of ZF can be derived
as $\delta H_{{\rm Z,e}}^{{\rm L}}=-(e/T_{{\rm e}})F_{0{\rm e}}\delta\phi_{{\rm Z}}$
and $\delta H_{{\rm Z,e}}^{{\rm NL}}=0$.

Ion responses to ZF can be obtained by implementing drift centre
transformation, i.e., $\delta H_{{\rm Z,i}}=\delta H_{{\rm dz,i}}\exp({\rm i}\hat{\lambda}_{{\rm dz}}\cos\theta)$,
with ${\hat\lambda}_{{\rm dz}}\equiv\hat{\omega}_{{\rm dri}}/\omega_{{\rm tr}}$
being the normalized drift orbit width. Substituting the expression of $\delta H_{{\rm Z,i}}$
to the gyrokinetic equation (\ref{eq:gke-zf}), noting the $\omega_{\rm Z}\ll v_{\rm ti}/(qR)$ ordering,  and taking the dominant flux surface
averaged quantities, the ion responses can be obtained as
\begin{eqnarray}
\overline{\delta H_{{\rm Z,i}}^{{\rm L}}}=\left|\theta_{{\rm Z}}\right|^{2}\frac{e}{T_{{\rm i}}}{\rm J}_{{\rm Z}}F_{0{\rm i}}\delta\phi_{{\rm Z}},
\end{eqnarray}
and
\begin{eqnarray}
\overline{\delta H_{{\rm Z,i}}^{{\rm NL}}} & = & \left|\theta_{{\rm Z}}\right|^{2}\dfrac{c}{B_{0}}k_{\theta n}\dfrac{e}{T_{{\rm i}}}{\rm J}_{n}^{2}F_{0{\rm i}}\dfrac{\omega_{*n{\rm i}}}{\omega_{n}^{2}}\partial_{r}|A_{n}|^{2},
\end{eqnarray}
where $\theta_{{\rm Z}}\equiv\overline{\exp(-{\rm i}\hat{\lambda}_{{\rm dz}}\cos\theta)}$
and  $\overline{\left(\cdots\right)}\equiv\int_{0}^{{\rm 2\pi}}\left(\cdots\right){\rm d}\theta/(2\pi)$  denotes  flux surface average. In deriving $\delta H^{\rm NL}_{\rm Z,i}$, the linear ion response to DW  given by Eq. (\ref{eq:DW_ion_linear}) is used. Furthermore, we have maintained only the meso- and macro-scale radial structures of ZF, averaging over the fine DW micro-scales assuming $\int_{-1}^{1}|\varPhi(nq-m)|^2{\rm d}(nq-m)=1$.
Substituting the particle responses of ZF to the charge quasi-neutrality
condition (\ref{eq:QN-condition}), the equation
for ZF nonlinear excitation can be readily obtained as
\begin{eqnarray}
\chi_{{\rm iZ}}\delta\phi_{{\rm Z}} & = & \dfrac{c}{B_{0}}k_{\theta n}\dfrac{\omega_{*n{\rm i}}}{\omega_{n}^{2}}\partial_{r}|A_{n}|^{2}.\label{eq:ZF_scalar}
\end{eqnarray}
Here, $\chi_{{\rm iZ}}\equiv1-\left\langle \left|\theta_{{\rm Z}}\right|^{2}{\rm J_{Z}^{2}}F_{0{\rm i}}/N\right\rangle \approx1.6k_{{\rm Z}}^{2}\rho_{{\rm ti}}^{2}q^{2}/\sqrt{\epsilon}$
  represents the neoclassical inertia enhancement \cite{MRosenbluthPRL1998}, where $\rho_{\rm ti}\equiv v_{\rm ti}/\omega_{\rm ci}$ is the ion Larmor radius defined by ion thermal velocity.
 Equation (\ref{eq:ZF_scalar}) gives a distinctive ZF temporal evolution with respect  to that of Ref. \cite{LChenPoP2000} (Note the $-{\rm i}\omega_{\rm Z}$ operator on the left hand side of Eq. (3) therein).  Equation (\ref{eq:ZF_scalar})
describes that   the ZF
grows at twice the instantaneous growth rate of DW, which is a typical
feature of the forced-driven process \cite{YTodoNF2010,ZQiuPoP2016,HWangPoP2020}, with the crucial role played by  the   nonlinear ion response to  ZF due to  thermal  ion nonuniformity. Meanwhile,
the forced driven process considered here is thresholdless, in contrast
to the spontaneous excitation of ZF by DWs via modulational instability, which requires sufficiently
large DWs amplitude to overcome the threshold due to  frequency mismatch \cite{LChenPoP2000}.
As a consequence, the forced-driven process is expected to occur  once DWs are driven
unstable, and that is the reason that forced-driven process
is universally observed in micro turbulence simulations; while, ZF
can be further excited via modulational instability after the amplitude of DWs
reaching certain threshold.

Performing inverse Fourier transformation, and integrating   in the radial direction,
Eq. (\ref{eq:ZF_scalar})  can be re-written  as
\begin{eqnarray}
\delta E_{\rm Z} & = & \dfrac{\sqrt{\epsilon}}{1.6q^{2}\rho_{{\rm ti}}^{2}}\dfrac{c}{B_{0}}k_{\theta n}\dfrac{\omega_{*n{\rm i}}}{\omega_{n}^{2}}\left|A_{n}\right|^{2},\label{eq:ZF-eq}
\end{eqnarray}
where $\delta E_{{\rm Z}}\equiv-{\rm \partial}_{r}\delta\phi_{{\rm Z}}$
is   the radial electric field of ZF.

\section{Nonlinear DW evolution due to forced driven ZF}\label{sec:feedback}

Next, we consider the feedback of the forced-driven ZF to the pump DW. The adiabatic electron response for
DW is adopted consistently with the $v_{\rm te}/(qR)\gg\omega_n$ ordering, i.e., $\delta H_{n{\rm ,e}}=0$. The ion responses to DW can be derived, assuming $|\omega_n|\gg|k_\parallel v_\parallel|$, as
\begin{eqnarray}
\delta H_{n,{\rm i}}^{{\rm L}}=\dfrac{e}{T_{\rm i}} F_{{\rm 0i}}\left(1-\dfrac{\omega_{*n{\rm i}}}{\omega_n}\right){\rm J}_{n}\delta\phi_{n},\label{eq:DW_ion_linear}
\end{eqnarray}
and
\begin{eqnarray*}
\delta H_{n{\rm ,i}}^{\rm NL} & = & \dfrac{c}{B_{0}}k_{\theta n}{\rm J}_{n}{\rm J_{Z}}\dfrac{e}{T_{{\rm i}}}F_{0{\rm i}}\dfrac{\omega_{*n{\rm i}}}{\omega_{n}^{2}}\delta\phi_{n}\delta E_{{\rm Z}}.
\end{eqnarray*}

Substituting the particle responses of DW to the charge quasi-neutrality
condition (\ref{eq:QN-condition}), the equation describing nonlinear
modulation of forced-driven ZF on DW can written as
\begin{eqnarray}
\epsilon_{0}\delta\phi_{n}+\dfrac{ck_{\theta n}}{B_{0}\omega_{n}}\delta\phi_{n}\delta E_{{\rm Z}} & = & 0,\label{eq:DW-eq}
\end{eqnarray}
where $\epsilon_{0}$ is the linear DWs dispersion relation. For the
proof of principle demonstration, the
electron DW in toroidal geometry is adopted for
later analysis, i.e., $\epsilon_{0}\delta\phi_{n}\rightarrow\hat{\epsilon}_{0}A_{n}\hat{\phi}\left(\eta\right)$,
with the  linear dielectric operator in the ballooning space  written as
\begin{eqnarray}
\hat{\epsilon}_{0} & = & \left(1-\dfrac{\omega_{*n{\rm i}}}{\omega_n}\right)\dfrac{c_{{\rm s}}^{2}}{r^{2}R^{2}}\partial_{\eta}^{2}+1-\dfrac{\omega_{*n{\rm e}}}{\omega_n}\nonumber \\
 &  & +\tau k_{\theta n}^{2}\rho_{{\rm ti}}^{2}\left[1+\hat{s}^{2}\left(\eta-\theta_{k}\right)^{2}\right]\nonumber \\
 &  & -\tau\left(1-\dfrac{\omega_{*n{\rm i}}}{\omega_n}\right)\dfrac{\omega_{{\rm di}}}{\omega_n}g\left(\eta,\theta_{k}\right),\label{eq:DW-DR}
\end{eqnarray}
and $\eta$ being  the extended poloidal angle along the equilibrium
magnetic field. Here, $\theta_{k}\equiv k_{r}/\left(n\partial q/\partial r\right)$, $c_{{\rm s}}\equiv\sqrt{2T_{{\rm e}}/m_{{\rm i}}}$
is the sound speed,  $\hat{\phi}(\eta)$ is the parallel mode structure
of DW,   $g\left(\eta,\theta_{k}\right)\equiv\cos\eta+\hat{s}\left(\eta-\theta_{k}\right)\sin\eta$
represents the  curvature, and  $\hat{s}\equiv r(\partial q/\partial r)/q$ is the magnetic shear. By combining
Eqs. (\ref{eq:ZF-eq}) and (\ref{eq:DW-eq}),
the equation describing the nonlinear evolution of DW via forced-driven ZF  modulation is given as
\begin{eqnarray}
\hat{\epsilon}_{0}A_{n}-\dfrac{\sqrt{\epsilon}}{1.6q^{2}\rho_{{\rm ti}}^{2}}\dfrac{\tau c^{2}k_{\theta n}^{2}}{B_{0}^{2}\omega_{n}^{2}}\left|A_{n}\right|^{2}A_{n} & = & 0,\label{eq:combined-eq1}
\end{eqnarray}
where we have assumed that DW is the ground state electron DW with $\omega_n\simeq\omega_0=\omega_{*n{\rm e}}/\left(1+k_{\theta n}^{2}\rho_{{\rm ti}}^{2}\right)$. Furthermore, we have integrated over the parallel mode structure, $\hat{\phi}(\eta)$, noting $\hat{\epsilon}_{0{\rm r}}\left(\omega_n=\omega_{0},r=r_{0},\theta_{k}=0\right)\hat{\phi}\left(\eta\right)=0$.
For radially localized fluctuation structures,
one can expand the DW eigenmode operator  $\hat{\epsilon}_{0}$ around $\theta_{k}=0$ and $r=r_{0}$ as
\begin{eqnarray}
\hat{\epsilon}_{0} & \approx & {\rm i}\left.\dfrac{\partial\hat{\epsilon}_{0{\rm r}}}{\partial\omega}\right|_{\omega_{0}}\left(\partial_{t}-\gamma_{{\rm L}}\right)+\dfrac{1}{2}\left.\dfrac{\partial^{2}\hat{\epsilon}_{0{\rm r}}}{\partial\theta_{k}^{2}}\right|_{0}\theta_{k}^{2}\nonumber \\
 &  & -\dfrac{\omega_{*n{\rm e}}(r)-\omega_{*n{\rm e}}(r_0)}{\omega_{0}}.\label{eq:DR-expansion}
\end{eqnarray}
Substituting Eq. (\ref{eq:DR-expansion}) into Eq. (\ref{eq:combined-eq1}),
it is found that the obtained nonlinear DW  equation is   a nonlinear
Schr\"odinger equation (NLSE), which can be explicitly written as
\begin{eqnarray}
\left(\partial_{t}-\gamma_{{\rm L}}+{\rm i}\tau\omega_{0}\rho_{{\rm ti}}^{2}\partial_{r}^{2}+{\rm i}\omega_{0}\varOmega(r)\right.\nonumber \\
\left.+{\rm i}\omega_{0}\dfrac{\alpha}{\tau}\dfrac{e^{2}}{T_{{\rm i}}^{2}}\left|A_{n}\right|^{2}\right)A_{n}(r,t) & = & 0,\label{eq:NLSE-1}
\end{eqnarray}
where $\alpha\equiv\left(\sqrt{\epsilon}/(1.6q^{2})\right)\left(k_{\theta n}^{2}\rho_{{\rm ti}}^{2}/4\right)\omega_{{\rm ci}}^{2}/\omega_{0}^{2}$
is the nonlinear coupling coefficient, and $\varOmega(r)=(\omega_{*n{\rm e}}(r)/\omega_{*n{\rm e}}(r_0)-1)$
represents the plasma nonuniformity. The second, third, and fifth
terms of Eq. (\ref{eq:NLSE-1}) represent, respectively,  the linear growth,
linear dispersiveness, and cubic nonlinearity introduced by the feedback
of forced driven ZF to DW. It is noteworthy that, the soliton generation due
to the balance of linear dispersiveness and nonlinear wave trapping  is an important topic, for its
potential relevance to DW turbulence spreading. In addition, plasma
nonuniformity at, e.g., pedestal region,  may  introduce a global potential,
 and prevent turbulence from spreading to linearly stable region \cite{ZQiuPoP2014}.
Due to the complexity of the nonlinear equation, especially  when
plasma nonuniformity is accounted for, the equation is mainly investigated numerically in the present  work.
For the convenience of numerical investigation, space and time are
normalized to ion Larmor radius $\rho_{{\rm ti}}$ and ion diamagnetic
frequency  at $r=r_0$ $\omega_{*n{\rm i}}(r_0)$, respectively, i.e., $r\rightarrow(r-r_0)/\rho_{{\rm ti}}$, $\gamma_{\rm L}\rightarrow\gamma_{\rm L}/\omega_{*n{\rm i}}$
$t\rightarrow\omega_{*n{\rm i}}t$, $k_{r}\rightarrow k_{r}\rho_{{\rm ti}}$ and $\omega\rightarrow\omega/\omega_{*n{\rm i}}$,
and $A_{n}$ is normalized to $e/T_{{\rm i}}$, i.e., $A=eA_{n}/T_{{\rm i}}$.
Then, the normalized form of Eq. (\ref{eq:NLSE-1}) can be written
as
\begin{eqnarray}
\left(\partial_{t}-\gamma_{{\rm L}}-{\rm i}\tau^{2}\partial_{r}^{2}-{\rm i}\alpha\left|A\right|^{2}-{\rm i}\tau\varOmega(r)\right)A & = & 0.\label{eq:NLSE}
\end{eqnarray}

Before proceeding with  numerical solution, it is necessary to derive
the conservation laws of the nonlinear system, which reveals
the essential information of the underlying physics and can be used to validate the numerical
results. The NLSE typically has two conservation laws, which are conservation
of ``mass'' (number of quasi-particles \cite{FZoncaPPCF2015}) and energy. The conservation of ``mass'' can be derived
by adding $A^{*}\times$Eq. (\ref{eq:NLSE}) to its complex conjugate,
which yields
\begin{eqnarray}
\partial_{t}|A|^{2}-{\rm i}\tau^{2}\partial_{r}\left(A^{*}\partial_{r}A-A\partial_{r}A^{*}\right) & = & 2\gamma_{{\rm L}}\left|A\right|^{2}.\label{eq:CL1}
\end{eqnarray}

Equation  (\ref{eq:CL1}) is a continuity equation with source on the
right hand side; while the second term on the left hand side represents
the divergence of the ``flux''   $J=A^{*}\partial_{r}A-A\partial_{r}A^{*}$.
Assuming vanishing flux at the boundary, the conserved quantity can
be readily obtained by integrating over the radial domain as
\begin{eqnarray}
\partial_{t}W & = & \left\langle 2\gamma_{{\rm L}}\left|A\right|^{2}\right\rangle _{r},\label{eq:CL-11}
\end{eqnarray}
where $\left\langle \cdots\right\rangle _{r}\equiv\int_{-\infty}^{+\infty}\cdots{\rm d}r$
is the integration over the whole radial domain and $W\equiv\left\langle \left|A^{2}\right|\right\rangle _{r}$
is the ``mass'' of the system. The right hand side of Eq. (\ref{eq:CL-11})
represents the source originates from the linear growth rate of DW.
Meanwhile, the energy conservation law can be derived by subtracting
$\partial_{t}A^{*}\times$Eq. (\ref{eq:NLSE}) from its complex conjugate,
which, following the same procedure, yields
\begin{eqnarray}
\partial_{t}E & = & 2\gamma_{\rm L}\left\langle A\partial_{t}A^{*}\right\rangle_{r}.\label{eq:CL2-1}
\end{eqnarray}

Here, $E\equiv\left\langle \tau^{2}\left|\partial_{r}A\right|^{2}+\alpha\left|A\right|^{4}+\tau\varOmega\left|A\right|^{2}\right\rangle _{r}$
is the total energy of the system, with three terms representing the
energy of DW, ZF, and the potential energy, respectively.

The numerical scheme used to solve the Eq. (\ref{eq:NLSE})
is the \textit{pseudospectral} method, which is characterized by multiplying
nonlinear terms in physical space and transforming back to Fourier
space instead of convolution sum in conventional spectral method.
More specifically, the NLSE is Fourier transformed in radial direction
into an ordinary differential equation, then, the forth order
Runge-Kutta method is applied to solve the temporal evolution equation.
The absorption boundary condition is applied to avoid un-physical
reflection back to the simulation domain, which is achieved by an
artificial damping layer near the boundary. Without loss of generality,
the nonuniformity is taken as $\varOmega(r)=\exp(-r^{2}/L_{{\rm p}}^{2})-1$, with $r=0$ corresponding to gradient steepening for increasing $r$, and
  $L_{{\rm p}}$ being the characteristic length of nonuniformity.
To estimate proper value of nonlinear coupling coefficient $\alpha$,
we have, for typical parameters of tokamaks, $\sqrt{\epsilon}/(1.6q^{2})\sim10^{-1}$,
$k_{\theta n}^{2}\rho_{{\rm ti}}^{2}/4\sim10^{-2}$, $\omega_{{\rm ci}}^{2}/\omega_{0}^{2}\sim10^{5}$,
which finally yields $\alpha\sim10^{2}$. So it is reasonable  to take
$\alpha=100$ in the following  numerical study, and $\tau=1$ is adopted. A convergence
study is also carried out based on the conservation of $W$ and $E$,
and it is found that accuracy converges at the number of grid
points $N_{\rm g}=512$ for $r\in[-100,100]$, in both uniform and nonuniform
cases. In the following numerical studies, the default grid setup
is $N_{\rm g}=1024$ for $r\in[-100,100]$.

\section{DW soliton generation in uniform plasmas \label{sec:Uniform-plasmas}}

In uniform plasmas, i.e., $\varOmega(r)=0$, the NLSE can be readily
solved using travelling wave transformation.  Furthermore,  the linear DW growth rate is turned off to focus on the  nonlinear evolution of a   DW with given amplitude.   Assuming a travelling
wave solution $A=\hat{A}(T,\xi)\exp(-{\rm i}\omega t+{\rm i}k_{r}r)$, where $\xi=r-v_{{\rm g}}t$
is the coordinate in the moving frame of the wavepacket, $T$ represents
the temporal evolution of the envelope in $\xi$ space, and $v_{{\rm g}}=2\tau^{2}k_{r}$
is the group velocity of the wavepacket, the NLSE (\ref{eq:NLSE}) can
then be written as an envelope equation
\begin{eqnarray}
\left(\partial_{T}-{\rm i}\tau^{2}\partial_{\xi}^{2}+{\rm i}(\tau^{2}k_{r}^{2}-\omega)-{\rm i}\alpha\hat{A}^{2}\right)\hat{A} & = & 0.\label{eq:uniform-travelling}
\end{eqnarray}

For the one-soliton solution with constant envelope, $\hat{A}$ is
stationary in $\xi$ space, i.e., $\partial_{T}\hat{A}=0$. The envelope
equation (\ref{eq:uniform-travelling}) can be reduced to
\begin{eqnarray*}
\left(\partial_{\xi}^{2}-\dfrac{\tau^{2}k_{r}^{2}-\omega}{\tau^{2}}+\dfrac{\alpha}{\tau^{2}}\hat{A}^{2}\right)\hat{A} & = & 0,
\end{eqnarray*}
which then yields the  one-soliton solution in hyperbolic secant function form as
\begin{eqnarray}
A_{S} & = & \sqrt{\dfrac{2}{\alpha}}{\rm sech}\left(r-v_{{\rm g}}t\right)e^{{\rm i}k_{r}r-{\rm i}\omega t}.
\end{eqnarray}

\begin{center}
\begin{figure}
\subfloat{\includegraphics[scale=0.3]{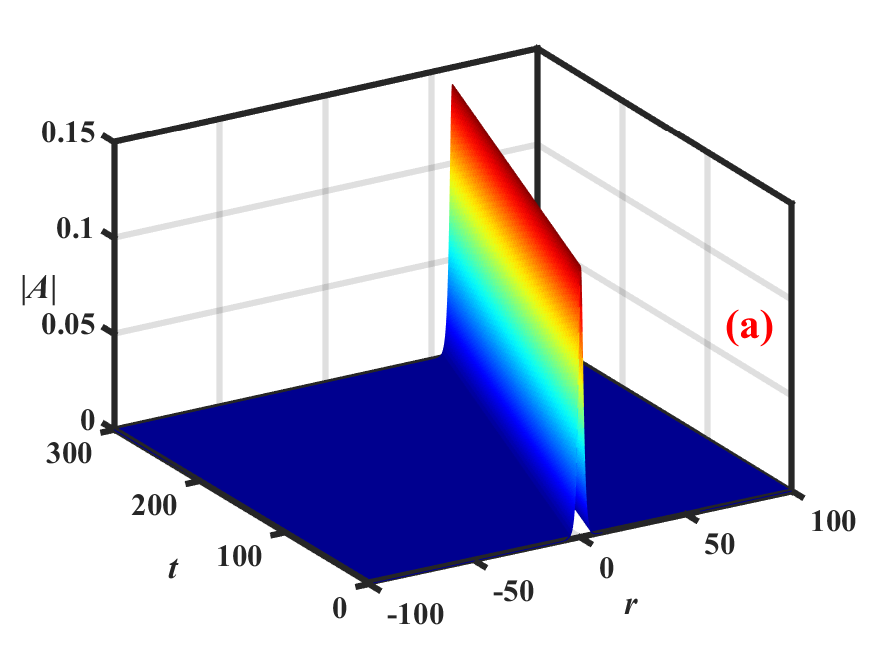}\label{Fig:one-soliton}}\subfloat{\includegraphics[scale=0.3]{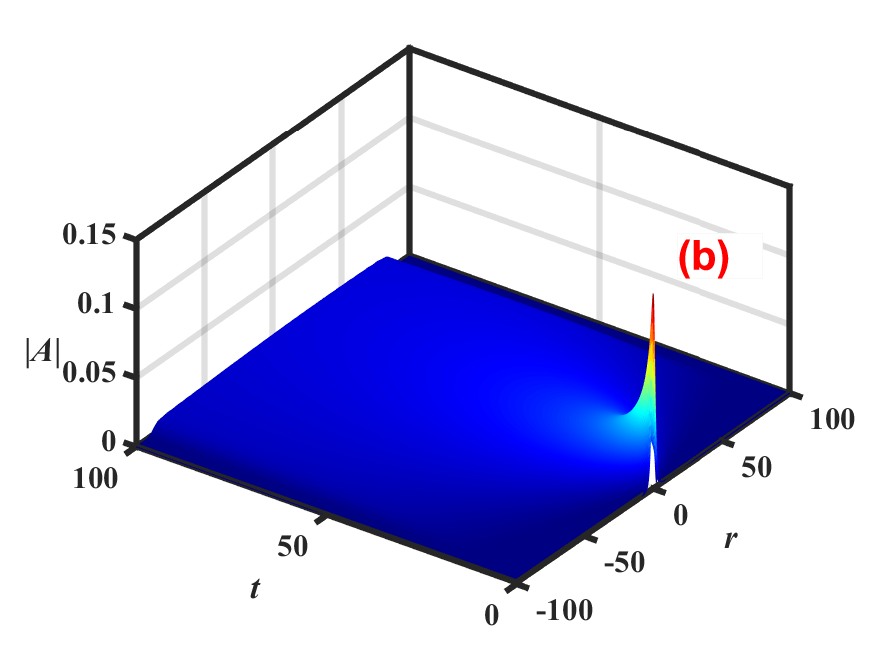}\label{Fig:one-soliton-L}}

\protect\caption{The  spatial-temporal evolution of the one-soliton solution (a) with and (b) without nonlinearity, where $k_r=0.1$ is used.}
\end{figure}

\par\end{center}

The one-soliton solution can be verified numerically, by taking $A_{S}(t=0)$
as initial condition, and the spatial-temporal evolution of DW is shown
in Fig. \ref{Fig:one-soliton}. It is observed that the initial envelope
propagates radially, with its shape and amplitude preserved. On the other hand,
in the absence of nonlinearity, the evolution equation of DW can be
derived from Eq. (\ref{eq:uniform-travelling}), which is a linear
Schr\"odinger (complex diffusion) equation
\begin{eqnarray}
\left(\partial_{T}-{\rm i}\tau^{2}\partial_{\xi}^{2}\right)\hat{A} & = & 0,\label{eq:diffusive}
\end{eqnarray}
which then describes the envelope of DW decays as dispersive wave packet $\hat{A}\propto1/\sqrt{T}\exp(-\xi^{2}/(\tau^2T))$,
i.e.,  the amplitude of DW decreases and its width becomes
wider to preserve the ``mass'' $W$, as shown in Fig. (\ref{Fig:one-soliton-L}). However, in the
presence of nonlinearity, the linear dispersiveness is balanced by
nonlinear trapping effect induced by the forced-driven ZF, and thus, soliton
solution can be established, as shown in Fig. \ref{Fig:one-soliton}.

\begin{center}
\begin{figure}
\subfloat{\includegraphics[scale=0.3]{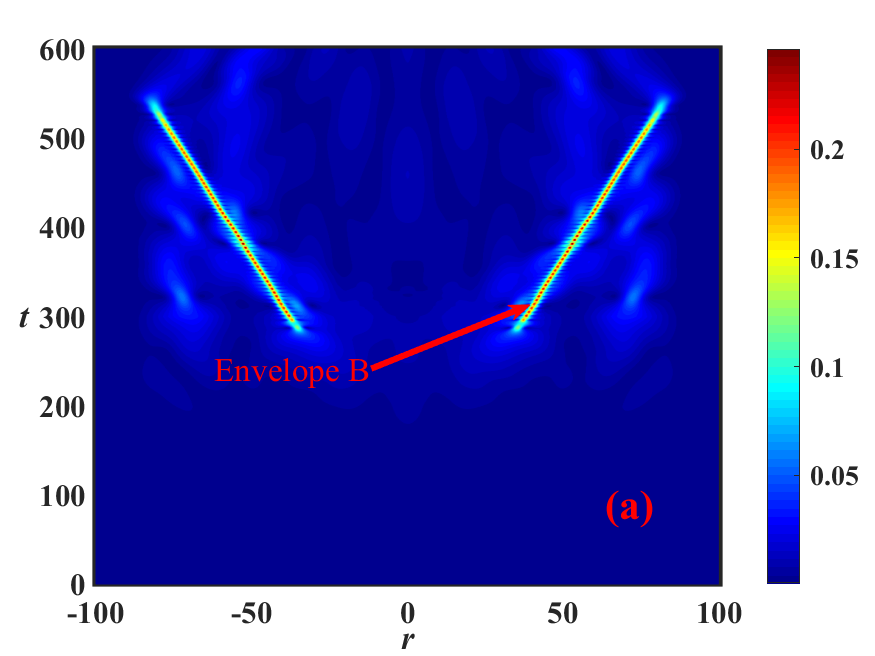}\label{Fig:soliton-noise}}\subfloat{\includegraphics[scale=0.3]{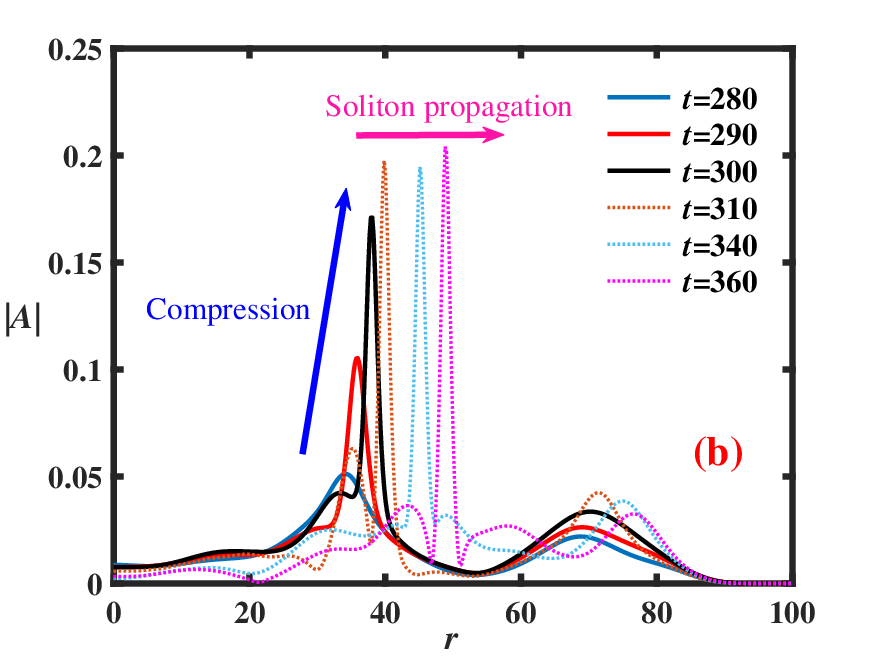}\label{Fig:noise-ms}}

\protect\caption{(a) The spatial-temporal evolution of the initial noise in uniform plasmas, with the red arrow pointing at the envelope ``B''. (b)
Snapshots of the radial mode structure of DW at a sequence of time, corresponding to the time in Fig. (a).}
\end{figure}

\par\end{center}

\begin{figure}
\includegraphics[scale=0.5]{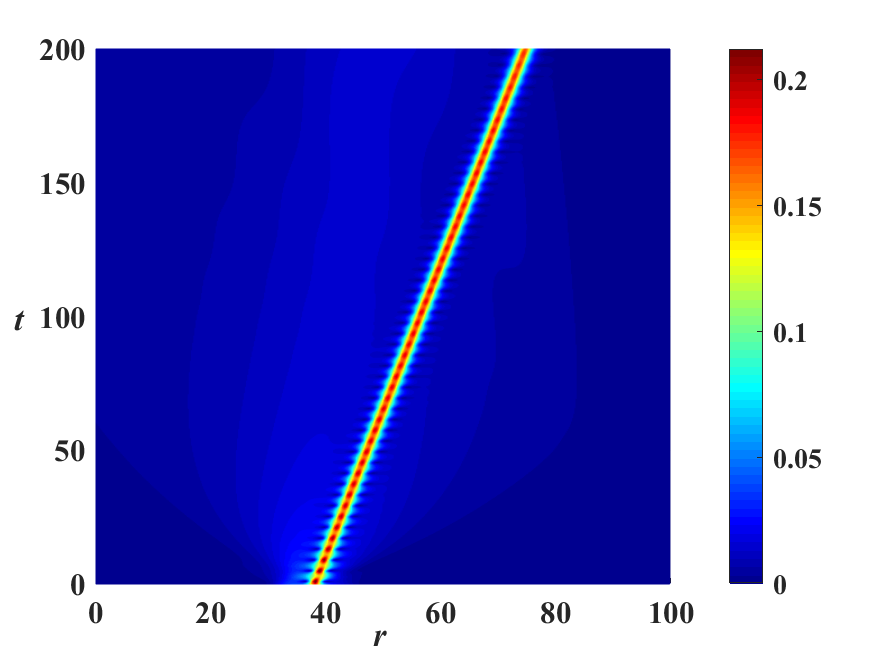}
\protect\caption{The spatial-temporal evolution of the envelope ``B'' of Fig. \ref{Fig:soliton-noise}.\label{fig:uniform_ic}}

\end{figure}

The forced-driven process may occur when DW amplitude is still small. Thus, it is
worthwhile to demonstrate the soliton formation due to ZF self-consistently
driven by   DW growing from noise level, in the presence of linear growth rate $\gamma_{\rm L}$.  In this case, DW is loaded  initially as   random noise,  which grows exponentially due to finite linear growth rate $\gamma_{{\rm L}}$; meanwhile,
the growth rate of zonal flow is $2\gamma_{{\rm L}}$. Since there
is little saturation mechanism for DW in the present model (no feedback to $\gamma_{\rm L}$ due to ZF scattering), the linear DW growth
rate is artificially turned off later to impose saturation.
Without loss of generality, $\gamma_{{\rm L}}$ is uniform in space,
i.e., $\gamma_{{\rm L}}=\gamma_{{\rm L}0}$ for $t\leq t_{{\rm c}}$,
and $\gamma_{{\rm L}}=0$ for $t>t_{{\rm c}}$, where $t_{{\rm c}}$
is the time of ``saturation''. The spatial-temporal evolution of DW is
shown in Fig. \ref{Fig:soliton-noise}, where $t_{{\rm c}}=300$,
$\gamma_{{\rm L0}}=0.025$ and $A_{0}=10^{-4}$ are adopted, with  $A_{0}$  being
the amplitude of initial noise. It is found that after its amplitude
reaches certain threshold at $t\simeq280$, the soliton structure
formation can be observed with a significant compression process,
which is more clearly shown in Fig. \ref{Fig:noise-ms}. In the
compression process from $t=280$ to $300$, the width of DW envelope
decreases and its amplitude grows significantly faster than the linear
growth rate, due to the self-trapping by forced-driven ZF and conservation of DW ``mass" $W$.
As the nonlinear trapping ($\propto|A|^{2}$) balances the dispersiveness,
the steady propagation of solitons can be established, as shown by
dashed curves in Fig. \ref{Fig:noise-ms}.  To
have a better view on the propagation of solitons, the envelope ``B''
labeled in Fig. \ref{Fig:soliton-noise} is isolated by multiplying
the mode structure at $t=310$ with a super-gaussian $\exp(-(r-r_{B})^{4}/L_{B}^{4})$,
where $r_{{\rm B}}$ is the position of the peak of ``B'', and $L_{{\rm B}}$
is the width of the filter.  The evolution of the envelope ``B''
is shown in Fig. \ref{fig:uniform_ic}, in which a soliton with periodically
oscillating amplitude can be observed. Thus, it is demonstrated
that in uniform plasmas,  linearly unstable  DW with a small initial amplitude can form soliton structures
after its amplitude reach certain threshold, with a significant compression.

The results shown above manifest the important feature of soliton
generation due to the forced-driven ZF, but it requires sufficiently
large DW amplitude. Thus, it is necessary to find out the threshold for soliton
formation to determine the relevance to realistic tokamak plasmas.  When dispersiveness dominates over the nonlinearity, DW
envelope decays with time as demonstrated by Eq. (\ref{eq:diffusive}). It is thus expected that as DW
amplitude increases, the decay rate of a given envelope should decrease
until it reaches steady state, and this defines the  threshold on DW amplitude  for soliton formation.
In   Fig. \ref{fig:decay}, the dependence of decay rate on DW amplitude
is given for different DW initial width $L_{{\rm d}}$, with
$A_0\exp(-r^{2}/L_{{\rm d}}^{2})$   given as initial condition  \footnote[40]{This expression is adopted  from Eq. (\ref{eq:weber_solution}) in nonuniform plasmas, with the characteristic mode width $L_{\rm d}$ determined by the nonuniformity scale length.}. It is found that even for the case with $L_{\rm d}=10$, the threshold on DW amplitude for steady state soliton formation, is about $e\delta\phi_n/T_{\rm i}\simeq 0.02$. Though strictly speaking, soliton structures
are characterized by preserving their  shape and amplitude during propagation,
 significant turbulence spreading can be observed when its
decay rate is smaller than characteristic time for turbulence spreading,
i.e. $L/v_{{\rm g}}$, where $L$ is the system size. As a consequence, the threshold on DW amplitude should be smaller than that corresponding to the vanishing decay rate.   The  typical DW fluctuation level expected  in tokamak experiments is $e\delta\phi_n/T_{{\rm i}}\simeq 0.05$, it is, thus, safe to conclude that the threshold on DW amplitude for soliton formation
is within the relevant range of experimental and simulation parameters,
and  may contribute to the turbulence spreading, as shown in
Fig. \ref{fig:turbulence_spreading}, where a DW envelope soliton is given as the initial condition.  It is found  that the initial DW perturbation flattens very quickly due to the linear dispersiveness in the linear case (L); however, in the nonlinear case (NL), the  soliton structure is well preserved, and can, thus, propagate into much broader radial extent than the linear case. Besides, it is also
worth mentioning  that the threshold for soliton formation decreases with increasing
$L_{{\rm d}}$, as shown in Fig. \ref{fig:decay},which is consistent with the result given by the inverse
scattering method. In fact, dispersiveness becomes
weaker for wider (smoother) gaussian envelope, thus, the DW amplitude needed
to nonlinearly balance the dispersiveness can be smaller. We  note that, strictly speaking,  the ``turbulence spreading''  in uniform plasma should be more precisely understood  as propagation of the initial DW perturbation, while the wave packet broadens due to  diffusion with $\Delta r\propto \sqrt{t}$, as there is no  ``linearly stable'' or ``linearly unstable'' regions in uniform plasmas.

\begin{center}
\begin{figure}
\includegraphics[scale=0.5]{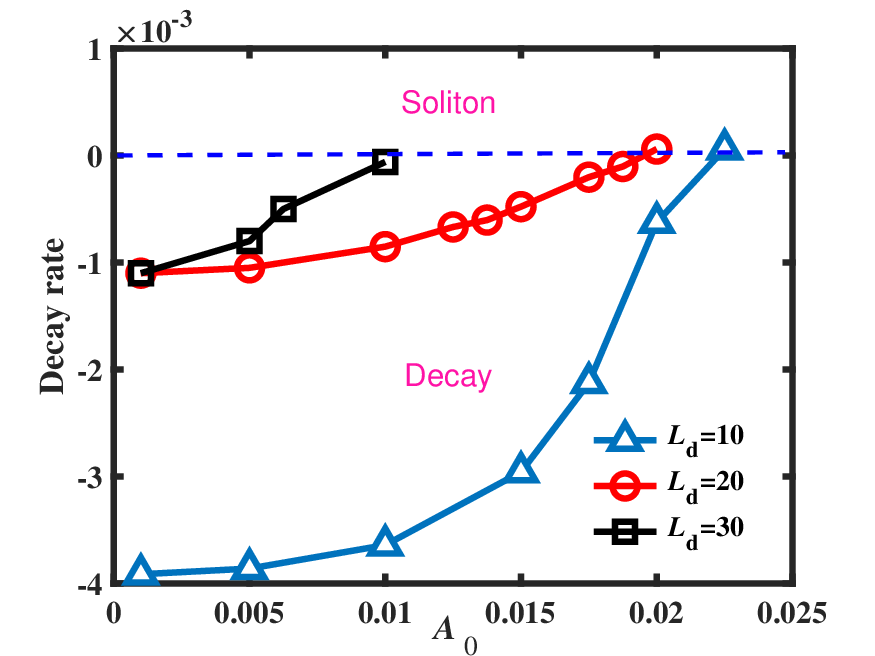}

\protect\caption{The dependence of decay rate of the envelope on its amplitude, with
the triangle, circular and square symbols representing that for the initial
width of envelope being 10, 20 and 30. The blue dashed line is the
separatrix of decaying envelope and solitons.\label{fig:decay}}
\end{figure}

\par\end{center}

\begin{center}
\begin{figure}
\includegraphics[scale=0.5]{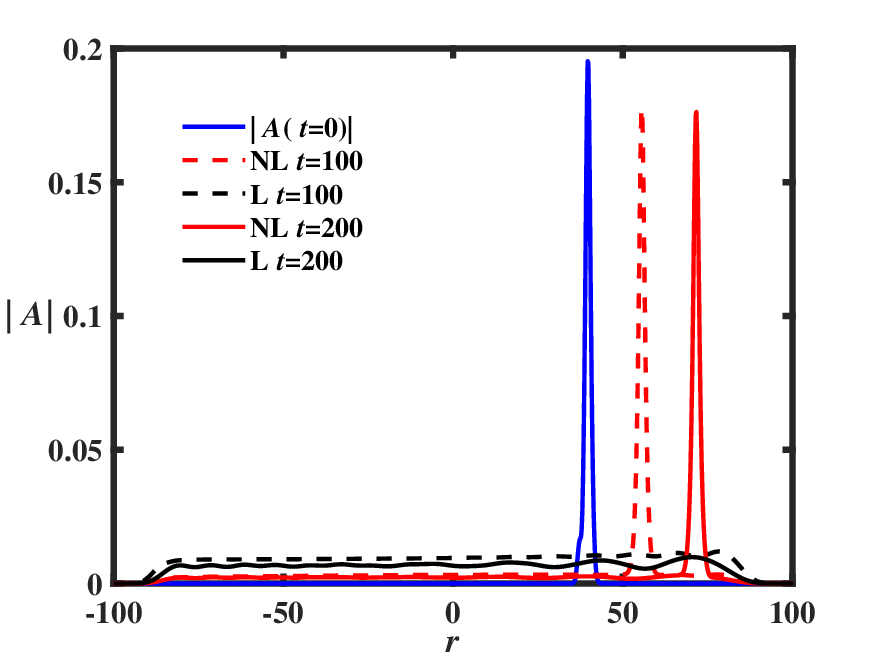}

\protect\caption{The radial mode structure in both linear and nonlinear cases at $t=100$ and $t=200$, with the envelope ``B''  of Fig. \ref{Fig:soliton-noise}  giving as initial condition. The blue solid curve is the envelope ``B'', the black and red dashed lines represent the linear and nonlinear mode structures at $t=100$, respectively, while, the black and red solid lines represent those at $t=200$.\label{fig:turbulence_spreading}}
\end{figure}

\par\end{center}

\section{DW soliton generation and propagation in nonuniform plasmas \label{sec:nonuniform-plasmas}}

In the last section, it is found  that in uniform plasmas   the formation of DW solitons can
be observed when linear dispersiveness is balanced by nonlinear
wave trapping effect induced by the  forced-driven ZF, which can contribute
to turbulence spreading. However, nonuniformity is intrinsic to magnetically
confined plasmas including the excitation of DWs. Consequently, it is particularly important to understand
the evolution of DW solitons in nonuniform plasmas. In the present analysis,
the plasma nonuniformity is introduced via the radial dependence of the diamagnetic drift frequency, $\varOmega(r)$, in Eq. (\ref{eq:NLSE}).
It is noteworthy  that, in Eq. (\ref{eq:NLSE}), the nonuniformity
not only serve as a nonuniform media for wave packet    propagation,
but also enables the formation of linear DW radial eigenstates before nonlinearity ($\propto|A|^2$) becomes significant. For   DWs with the amplitude well below the threshold for soliton formation, Eq. (\ref{eq:NLSE}), in the linear limit, reduces to
\begin{eqnarray}
\left(\partial_{r}^{2}+\dfrac{\omega}{\tau^{2}}+\dfrac{1}{\tau}\varOmega(r)\right)A & = & 0.\label{eq:NLSE-potential}
\end{eqnarray}

Here, $\omega={\rm i}(\partial_t-\gamma_{\rm L})$ is the linear eigenfrequency. For the gaussian-shape nonuniformity considered
in this work \footnote[41]{This  is  a reasonable assumption as one expands $\varOmega(r)$ around its local extreme.}, i.e., $\varOmega(r)=\exp(-r^{2}/L_{{\rm p}}^{2})-1$, expanding about $r=0$, Eq. (\ref{eq:NLSE-potential}) becomes approximately the well-known Weber
equation. The corresponding eigenfrequency and eigenfunctions are given by
\begin{eqnarray}
A=A_{0}\exp(-r^{2}/(2L_{{\rm d}}^{2})){\rm H}_{\ell}(r/L_{{\rm d}}),\label{eq:weber_solution}
\end{eqnarray}
and $\omega=\tau^{3/2}(2n+1)/L_{\rm p}$, respectively, where $L_{{\rm d}}=(\tau L_{{\rm p}}^{2})^{1/4}$, and $H_{\ell}$ is the
$\ell$-th Hermite polynomial. Taking the lowest eigenstate as initial condition, Fig. \ref{Fig:eigenstate_L} demonstrates that the linearized Eq. (\ref{eq:NLSE}) does retain the linear eigenmode solution. However, when the amplitude of DW increases  and nonlinearity becomes more significant, the linear eigenstates
might be qualitatively modified, as shown in Fig. \ref{Fig:eigenstate_NL}.

\begin{center}
\begin{figure}
\subfloat{\includegraphics[scale=0.3]{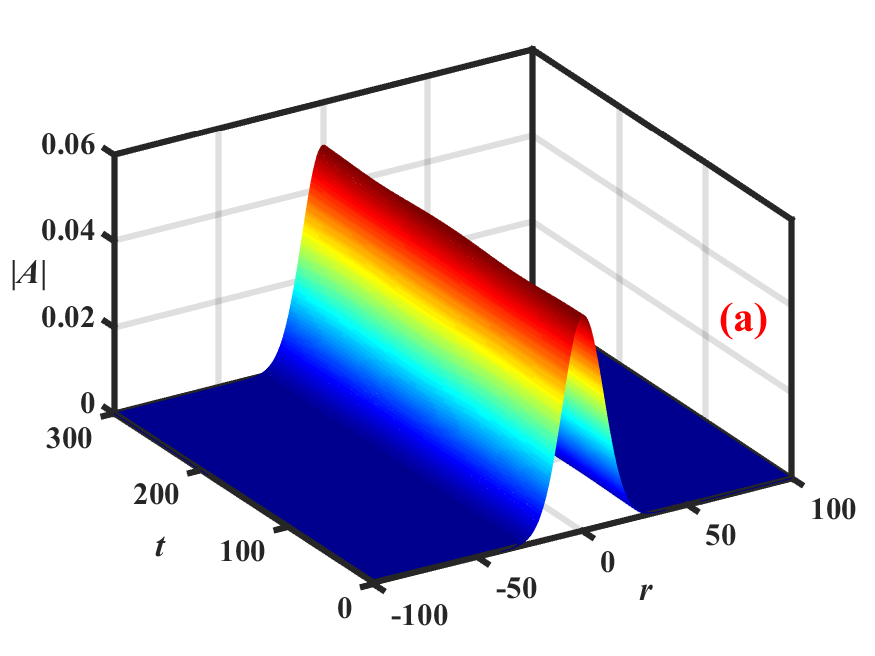}\label{Fig:eigenstate_L}}\subfloat{\includegraphics[scale=0.3]{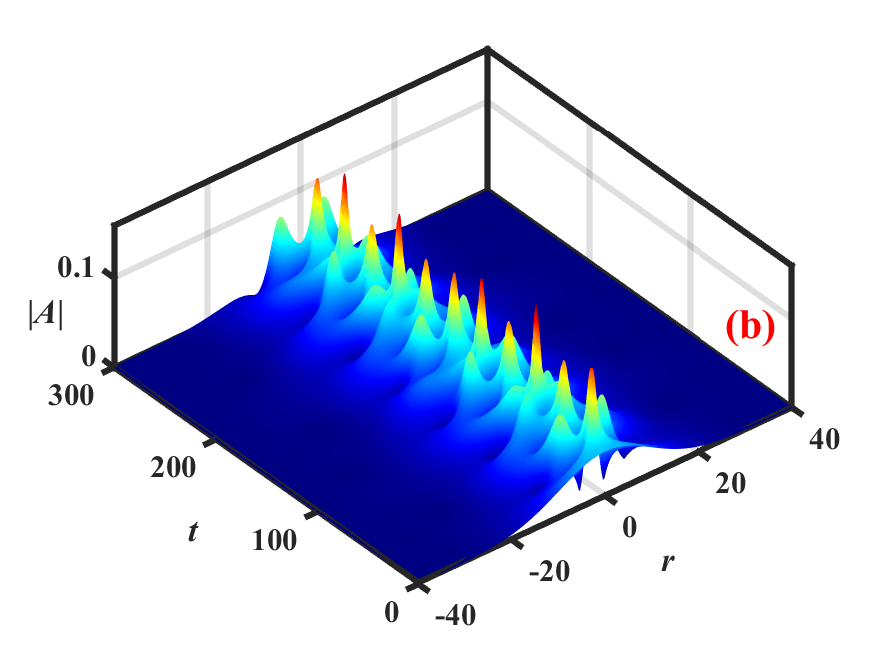}\label{Fig:eigenstate_NL}}
\protect\caption{The spatial-temporal evolution of the lowest eigenstate ($\ell=0$) of the nonuniform plasma (a) without and (b) with nonlinearity. Here, $L_{\rm p}=100$, $A_0=0.05$ are used.}
\end{figure}
\par\end{center}

Among various potential effects of the nonlinearity on DW envelope evolution, it is particularly important to investigate if and how nonlinearities affect the trajectory of the wavepacket, which could qualitatively determine the
radial extent of turbulence spreading.
To demonstrate the propagation of solitons,   the spatial-temporal evolution
of the fourth eigenstate is shown in Fig. \ref{Fig:2nd_DW},
in which propagation and collisions of solitons can be clearly observed.
Since trajectories of these solitons overlap with each other, the turning
points for each of them is difficult to determine. In order to observe
the propagation of a single soliton, soliton ``C'' labeled
in Fig. \ref{Fig:2nd_DW} is, again, isolated by a filter, following the same
procedures as that in former section. The spatial-temporal evolution
of the soliton ``C'' is shown in Fig. \ref{Fig:2nd_DW_ic}, where the soliton is reflected back and forth between   the turning
points located at $\pm r_{\rm tp}$, where the packet velocity vanishes.

\begin{center}
\begin{figure}
\subfloat{\includegraphics[scale=0.3]{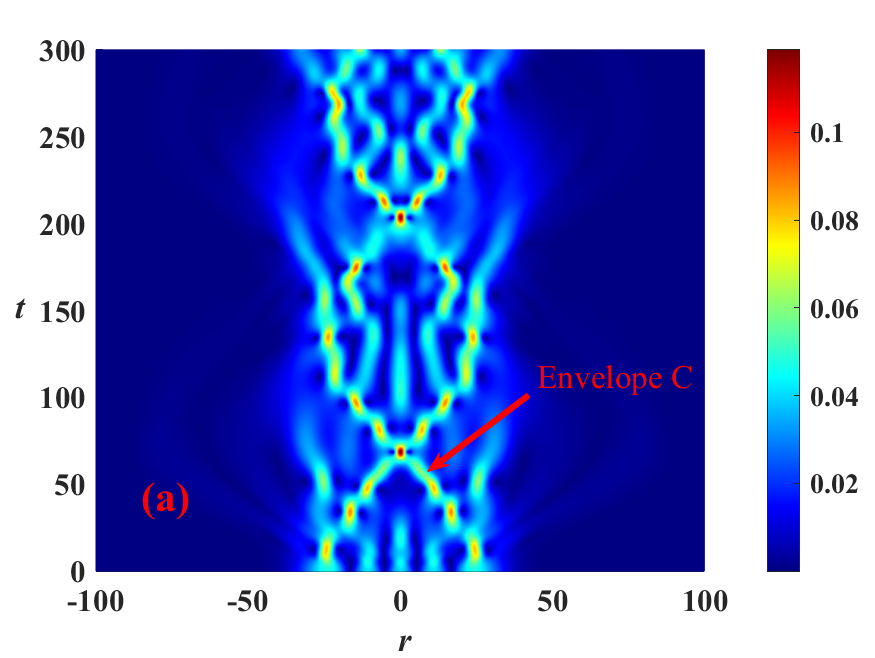}\label{Fig:2nd_DW}}\subfloat{\includegraphics[scale=0.3]{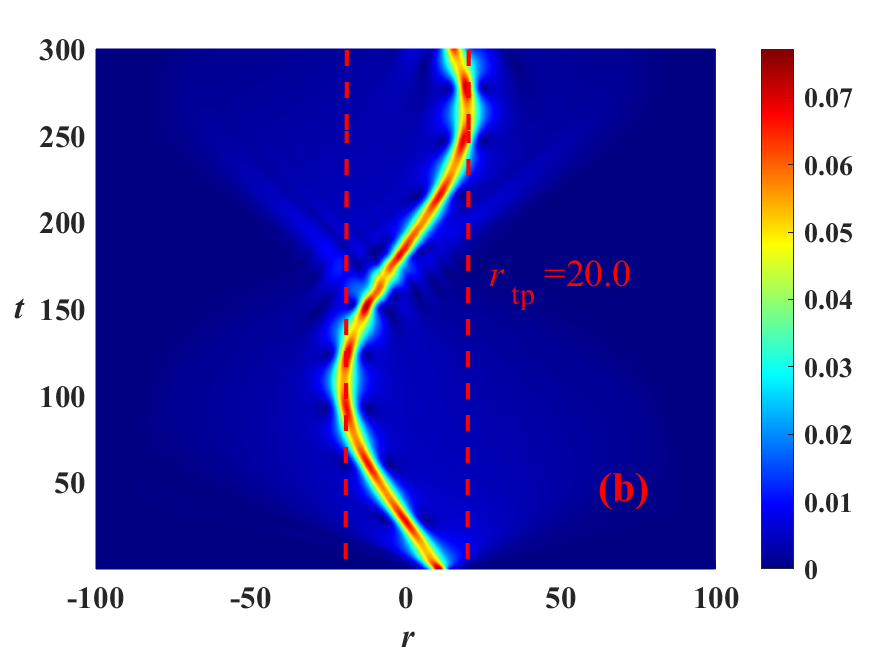}\label{Fig:2nd_DW_ic}}

\protect\caption{(a) The spatial-temporal evolution of the fourth order eigenstate, with the red arrow pointing at the envelope ``C''. Here, $L_{{\rm p}}=100$ and $A_0=0.05$. (b) The spatial-temporal evolution
of the envelope ``C'' alone, whose extent of propagation is bounded
inside turning points indicated by two red dashed lines.}
\end{figure}

\par\end{center}

By artificially changing  the amplitude of the envelope ``C'', the
dependence of  radial position of the turning point  $r_{{\rm tp}}$ on DW amplitude (i.e., nonlinearity)  can be
obtained and shown in Fig. \ref{Fig:rtp_A}. It is found that, within realistically reasonable DW amplitudes, $r_{{\rm tp}}$
is not sensitive to the strength of the nonlinearity, i.e., the radial
extent for  DW propagation  is nearly the same as in the linear 
case. In the linear limit, the WKB expression is given by
Eq. (\ref{eq:NLSE-potential}), i.e.,
\begin{eqnarray}
\omega+\tau\varOmega(r) & = & \tau^{2}k_{r}^{2},\label{eq:WKB}
\end{eqnarray}
and turning points are located at $\tau\varOmega(r_{\rm tp})=\omega$. Assuming
a constant $\omega$, we obtain
$r_{\rm tp}$ as
\begin{eqnarray}
r_{{\rm tp}} & = & \pm L_{\rm p}\sqrt{-\ln\left(1-\tau k_{r0}^{2}\right)}.\label{eq:rtp}
\end{eqnarray}
Here, $k_{r0}=\omega/\tau^2$ is the radial wavenumber at $r=0$. It is obvious that $r_{{\rm tp}}$ is proportional to $L_{{\rm p}}$,
and increases with increasing $k_{r0}$. The expression for $r_{{\rm tp}}$ given by Eq.
(\ref{eq:rtp}) can be examined numerically,  and the
dependence of $r_{{\rm tp}}$ on $k_{r0}$ is  shown by  the
blue solid and dashed curves in Fig. \ref{Fig:rtp-kr}, where  theoretical
and numerical values agree well with each other.

\begin{center}
\begin{figure}
\includegraphics[scale=0.5]{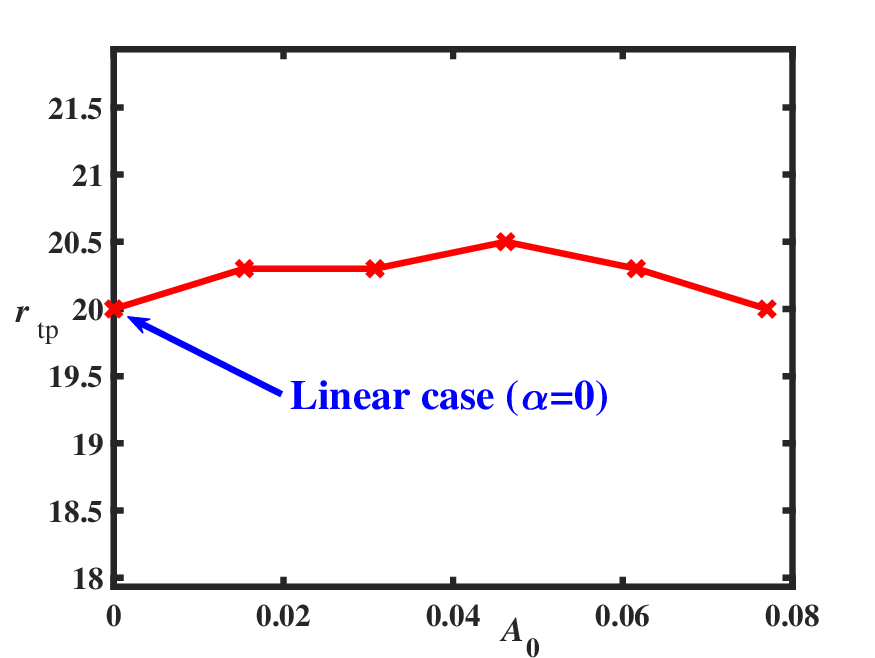}\protect\caption{The dependence of $r_{{\rm tp}}$ on the amplitude of soliton ``C'',
which is adjusted by multiplying the original envelope with ratios.\label{Fig:rtp_A}}
\end{figure}

\par\end{center}

That the wavepacket propagates, in the lowest order, as in the linear case, suggests taking the envelope  as $A=\hat{A}(T,\xi)\exp({\rm i}\int^r k_{r}{\rm d}r'-{\rm i}\omega t)$, with $k_r(r)$ satisfying the linear Eq. (\ref{eq:WKB}).
 The NLSE Eq. (\ref{eq:NLSE}) then becomes
\begin{eqnarray}
\left[\partial_{T} -{\rm i}\tau^{2}\partial_{\xi}^{2} -{\rm i} \alpha\hat{A}^{2} + \tau^2\partial_rk_r \right. \nonumber \\
\left.-{\rm i}\left(\omega+\tau\varOmega(r)-\tau^{2}k_{r}^{2}\right)\right]\hat{A} & = & 0,\label{eq:envelope}
\end{eqnarray}
which, for regions far from the turning points, can be solved perturbatively by expanding $k_r$ as $k_r=k_0+k_1+k_2+\cdots$, with $k_0\gg k_1\gg k_2\gg\cdots$. Truncating the solution at the first order, the solution is $\hat{A}\propto 1/\sqrt{k_0}\exp({\rm i}\int^r k_0{\rm d}r')$, in which $k_0=\sqrt{\omega/\tau^2+\varOmega(r)/\tau}$, if it is consistently assumed that the amplitude is sufficiently small that it enters at second order in the perturbation expansion. It implies that, as the wavepacket propagates away from local minimum of the potential well, its amplitude can be amplified due to the radial variation of the group velocity, i.e., the fourth term in the left hand side of Eq. (\ref{eq:envelope}), and vice versa. While, the trajectory of the envelope, $r_{\rm p}=\int^t v_{\rm g}{\rm d}t'$, is determined by $k_0$, which is not affected by the nonlinearity in the weak amplitude expansion defined above, consistent with Fig. \ref{Fig:rtp_A}.

The WKB solution breaks down near the turning points, in which the potential well can be expanded around the turning points, and Eq. (\ref{eq:envelope}) can be reduced to 
\begin{eqnarray}
(\partial_{T}-{\rm i}\tau^{2}\partial_{\xi}^{2}-{\rm i}\alpha\hat{A}^{2}-{\rm i}\tau\dfrac{\partial\varOmega}{\partial r}\Delta r)\hat{A} & = & 0.\label{eq:NLSE-nonuniform}
\end{eqnarray}
Here, $\Delta r=r-r_{\rm p}$ is the small deviation from the turning points. Considering $\partial_\xi^2\sim1/\Delta^2\sim|\hat{A}|^2\gg\Delta r$ ordering for solitons, with $\Delta$ being the width of the wavepacket, Eq. (\ref{eq:NLSE-nonuniform}) is essentially the NLSE in the moving
frame of the wavepacket. This implies that the evolution of the envelope
is governed by the same equation as that in uniform plasmas, except for the reflection induced by external potential $\partial\varOmega/\partial r\Delta r$, which can be neglected if the width of the wave packet is smaller than the distance from the turning point.
Above all, it is found that the trajectory and turning
points of the envelope are determined by the system nonuniformity via $\omega+\tau\varOmega(r) =\tau^2 k_r^2$, i.e., Eq. (\ref{eq:WKB}).
Thus, in nonuniform plasmas, the nonlinearity  serves as a local
potential well to balance the dispersiveness, while  the system nonuniformity serves as  a global potential well, which introduces extra trapping effect to the envelope and essentially determines the
trajectory  for envelope propagation. As a consequence, in  nonuniform plasmas, the threshold on DW amplitude for soliton formation could  be lower than that in uniform plasmas.

There is another aspect to demonstrate the extra trapping effect of
DW by the nonuniformity. The bounce time, which is the period  required for the envelope to travel  between two turning points $r=-r_{{\rm tp}}$ and $r=r_{{\rm tp}}$,
is given by
\begin{eqnarray}
\tau_{{\rm b}} & = & 2\int_{0}^{r_{{\rm tp}}}\dfrac{{\rm d}r}{v_{{\rm g}}(r)},\label{eq:taub}
\end{eqnarray}
where $v_{{\rm g}}(r)=2\tau^{2}\sqrt{k_{r0}^{2}-(1-\exp(-r^{2}/L_{{\rm p}}^{2}))}$
is the group velocity of the envelope. The dependence of $\tau_{{\rm b}}$
on $k_{r0}$ is shown by the  red solid curve  in Fig. \ref{Fig:rtp-kr}.
It is found that for modes with $k_{r0}\leq0.5$, after being reflected
by the turning points, these  components with different $k_{r0}$ will meet at $r_0$ at almost the same time, though they
have different group velocities. The evolution of the envelope
``C'' in the absence of nonlinearity is shown in Fig. \ref{Fig:2nd_DW_ic_L},
which can be compared with its nonlinear counterpart in Fig. \ref{Fig:2nd_DW_ic}.
It is observed in Fig. \ref{Fig:2nd_DW_ic_L} that linear dispersiveness
dominates the mode dynamics  in the beginning,   then peaks emerges with time  intervals   $\varDelta t$ being about  162, which is the $\tau_{{\rm b}}$
corresponding to the envelope ``C'' with $k_{r0}\leq 0.5$. Thus, the linear potential well introduced
by the system nonuniformity has extra trapping effect to the envelopes in the absence of nonliearity,
even though the envelope ``C'' is not a linear radial eigenstate of the system. In fact, it can be observed that
the amplitude of peaks decreases each time they emerge, because high-$k_{r}$
modes have  larger bounce time, thus, they will not collide
at the same time. Here, it is worthwhile drawing a similarity of this phenomenon with the spatial bunching of beam electrons in the beam plasma system \cite{TOneilPoF1971}. Even in that case, in fact, the spatial bunching, which is the counterpart of the amplitude peaking discussed here, is consequence of the ``isochronism'' in the particle (wave-packet) oscillations.
 Above all, in  nonuniform plasmas, the soliton
forms as it is trapped by the localized potential well induced by cubic
nonlinearity. Then the soliton is reflected by the global
potential well given by  system nonuniformity, suppressing the DW turbulence spreading into linearly stable region via soliton generation.

\begin{center}
\begin{figure}
\subfloat{\includegraphics[scale=0.3]{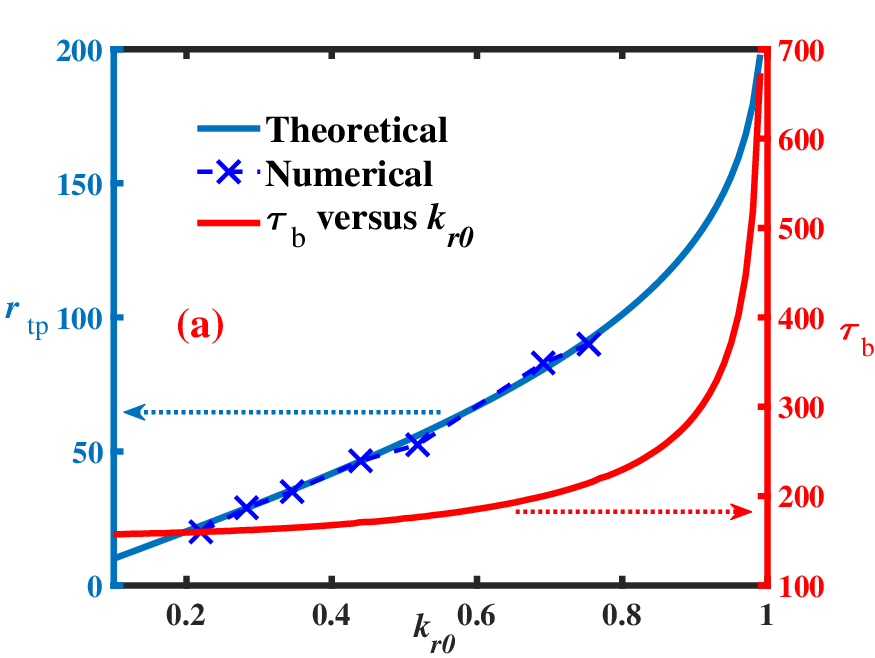}\label{Fig:rtp-kr}}\subfloat{\includegraphics[scale=0.3]{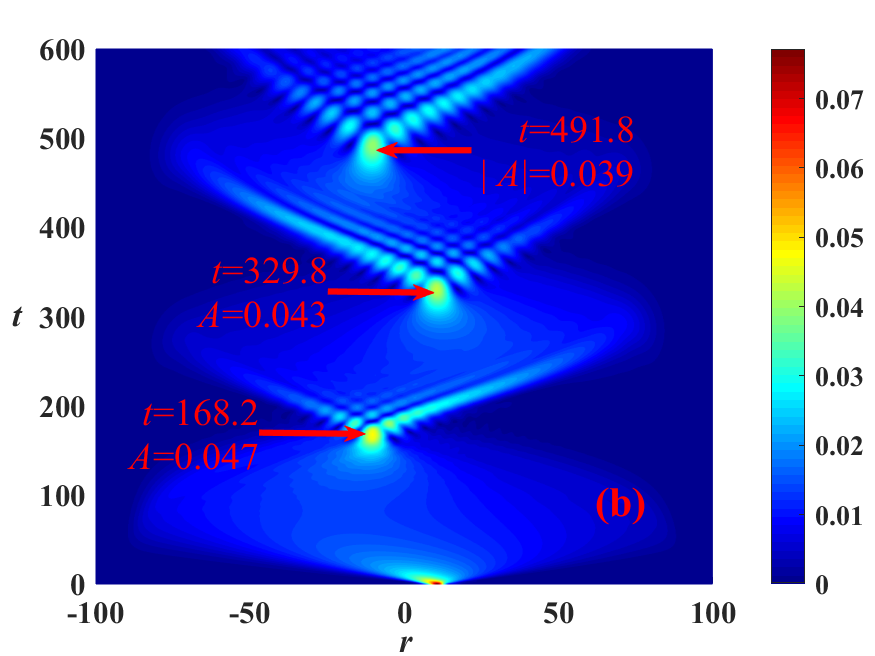}\label{Fig:2nd_DW_ic_L}}

\protect\caption{(a) The dependence of $r_{{\rm tp}}$ and $\tau_{{\rm b}}$ on $k_{r0}$,
in which the blue solid curve and blue crosses are the theoretical and numerical values of $r_{\rm tp}$, respectively, corresponding to the vertical
axis on the left; while, the red solid  curve respresents  $\tau_{\rm b}$ given by Eq. (\ref{eq:taub}), corresponding to the vertical
axis on
the right. (b) The evolution of the envelope ``C'' alone in
the absence of nonlinearity, with the red arrows representing the
time and amplitude of the corresponding peaks.}
\end{figure}

\par\end{center}

\section{Conclusion and discussion\label{sec:Conclusion-and-discussion}}

In this work, a paradigm model of drift waves (DWs) self regulation via the forced driven
 zonal flow (ZF) is derived using nonlinear
gyrokinetic theory. The obtained nonlinear DW equation is a nonlinear
Schr\"odinger equation (NLSE), in which the linear dispersiveness, linear
growth rate, plasma nonuniformity and cubic nonlinearity induced by
feedback of forced driven ZF to DW are self-consistently included.
The NLSE is systematically investigated in both  uniform and nonuniform
plasmas.

In uniform plasmas, soliton structures     can form as  DW amplitude reaching
the  threshold  for  the  cubic nonlinearity to balance  the linear
dispersiveness; and,  lead to  turbulence spreading via convective DW soliton propagation. The
threshold for soliton formation is found  to be  $e\delta\phi_n/T_{{\rm i}}\simeq0.02$,
well  within the experimentally relevant parameter regime. As such
  forced driven generation of  ZF by DW turbulence  is universally observed  in numerical simulations,
it is of interest to further investigate the soliton formation
in simulations.

In nonuniform plasmas, the evolution of the corresponding linear radial eigenstates
is investigated. It is found that the extent for wave propagation
is not sensitive to either the existence or strength of the nonlinearity, so  in
nonuniform plasmas, solitons can not extend beyond the range bounded
by the turning points induced by plasma nonuniformity.   As a result, in realistic geometry with intrinsic plasma nonuniformity, DW solitons can indeed form, however, it doesn't further extend  turbulence spreading to linearly stable region.  The plasma nonuniformity, however, can slightly reduce the threshold on DW amplitude required for soliton generation, due to the additional trapping by the potential well introduced by diamagnetic drift frequency or any other nonuniformity.

An important theoretical progress  of this work is the gyrokinetic  description   of forced-driven excitation of ZF by DWs,   commonly
observed in numerical simulations, which is a significant component
of DW-ZF interactions.  It is found that forced-driven excitation of ZF is through the nonlinear ion response to ZF induced by plasma nonuniformity, in contrast to the radial envelope modulation for spontaneous  excitation \cite{LChenPoP2000}.  This mechanism is, in fact, shown to be universal, and has been recently discussed in Ref. \cite{MFalessiNJP2023}, where the concept of zonal state (ZS) \cite{MFalessiPoP2019,FZoncaJPCS2021} was introduced as the self-consistent plasma equilibrium that is formed due to the excitation of ZF by plasma self-interactions and of their corresponding counterpart in the phase space, i.e., the phase space zonal structures (PSZS) \cite{FZoncaNJP2015}. Specific applications are given, e.g., in the ZF forced driven by toroidal Alfv\'en eigenmode  in nonuniform plasmas \cite{LChenWLIS2023} to investigate the effects of ZF on Alfv\'en eigenmode nonlinear saturation and indirect effects on DW stability via forced-driven ZF mediation. Further analyses of these nonlinear interactions leading to self-consistent  structure formation and corresponding results will be presented in future publications.

\section*{Data availability}

The data that support the findings of this study are
available from the corresponding author upon reasonable
request.

\section*{Acknowledgement}
This work was  supported by  the National Science Foundation of China under Grant Nos. 12275236 and 12261131622, and  Italian Ministry for Foreign Affairs and International Cooperation Project under Grant  No. CN23GR02.
 This work was also supported by the EUROfusion Consortium, funded by the European Union via the Euratom Research and Training Programme (Grant Agreement No. 101052200 EUROfusion). The views and opinions expressed are, however, those of the author(s) only and do not necessarily reflect those of the European Union or the European Commission. Neither the European Union nor the European Commission can be held responsible for them.

\end{document}